\documentclass[prl,aps,twocolumn,amsfonts]{revtex4}

\usepackage{graphicx}
\usepackage{epsfig}

\def\qed{\leavevmode\unskip\penalty9999 \hbox{}\nobreak\hfill
     \quad\hbox{\leavevmode  \hbox to.77778em{%
               \hfil\vrule   \vbox to.675em%
               {\hrule width.6em\vfil\hrule}\vrule\hfil}}
     \par\vskip3pt}

\def\ibb #1{\leavevmode\hbox{\kern.3em\vrule
     height 1.5ex depth -.1ex width .4pt\kern-.3em\rm#1}}

\newcommand{\be}[1]{\begin{equation} #1 \end{equation}}
\newcommand{\bea}[1]{\begin{eqnarray} #1 \end{eqnarray} }
\newcommand{\ba}[2]{\left(\begin{array}{#1}#2\end{array}\right)}

\newtheorem{theorem}{Theorem}

\newtheorem{lemma}{Lemma}

\begin{document}

\title{Normal forms and entanglement measures for multipartite quantum states}
\author{Frank Verstraete, Jeroen Dehaene and Bart De Moor\\
Katholieke Universiteit Leuven,
Department of Electrical Engineering, Research Group SISTA\\
Kasteelpark Arenberg 10, B-3001 Leuven, Belgium }

\date{August 6, 2001}

\begin{abstract}
A general mathematical framework is presented to describe local
equivalence classes of multipartite quantum states under the
action of local unitary and local filtering operations. This
yields multipartite generalizations of the singular value
decomposition. The analysis naturally leads to the introduction of
entanglement measures quantifying the multipartite entanglement
(as generalizations of the concurrence and the 3-tangle), and the
optimal local filtering operations maximizing these entanglement
monotones are obtained. Moreover a natural extension of the
definition of GHZ-states to e.g. $2\times 2\times N$ systems is
obtained.
\end{abstract}
 \maketitle

One of the major challenges in the field of quantum information
theory is to get a deep understanding of how local operations
assisted by classical communication (LOCC) performed on a
multipartite quantum system can affect the entanglement between
the spatially separated systems. In this paper we investigate this
problem in the case that only operations on one copy of the system
are allowed. This is different from the general setup of
entanglement distillation, where global operations on a large
(infinite) number of copies are performed to concentrate the
entanglement in a few copies. The main motivation of this work was
to characterize the optimal filtering operations (SLOCC) to be
performed on one copy of a multipartite system such that, with a
non zero chance, a state with maximal possible entanglement is
obtained. In other words, we want to design the optimal filtering
operations for a given state, such that with a certain chance we
prepare the optimal attainable one. Of course this leads to the
introduction of local equivalence classes.

In the case of a pure state of two qubits, this optimal filtering
procedure is commonly known as the Procrustean method
\cite{BBP96a}. Following the work of Gisin \cite{Gi96}, Horodecki
\cite{HHH97},Linden et al. \cite{LMP98} and Kent et al.
\cite{Ken98,KLM99}, the optimal filtering procedure for mixed
states of two qubits was recently derived in \cite{VDD01b}. In
this paper we extend these ideas to pure and mixed multipartite
systems of qudits of arbitrary dimension.

The optimal filtering operations in \cite{VDD01b} were derived by
proving the existence of a decomposition of a mixed state of two
qubits as a unique Bell diagonal state multiplied left and right
by a tensor product representing local operations. A Bell diagonal
state is special in the sense that one party alone cannot acquire
any information at all about the state: its local density operator
is equal to the identity. This can readily be generalized to
multipartite systems of arbitrary dimensions, and the existence of
local operations transforming a generic state to a unique state
with all local density operators equal to the identity will be
proved. In the case of pure states, this decomposition leads to a
transparent method of deriving essentially different states such
as GHZ- and W-states \cite{DVC00}.

We then proceed to show that all quantities exhibiting some kind
of invariance under the considered SLOCC operations are
entanglement monotones \cite{Vid00}. It is shown that the
concurrence and the 3-tangle, introduced by Wootters et al.
\cite{Woo98,CKW00}, belong to this class of entanglement measures.
Therefore a natural generalization of these measures is obtained
to systems of arbitrary dimensions and an arbitrary number of
parties.

A subsequent part of the paper is concerned with finding the
optimal filtering operations for a given multipartite state. It is
shown that the SLOCC operations bringing a state into its unique
normal form maximize all the introduced entanglement monotones.
This was expected in the light of the work by Nielsen about
majorization \cite{Nie99}: the notion of local disorder is
intimately connected to the existence of entanglement.

Finally, the appendix presents some results on the
characterization of local unitary equivalence classes, yielding a
natural and constructive but non-unique generalization of the
singular value decomposition to the multilinear setting.

\subsection{Normal forms under SLOCC operations}

Let us first consider the case of pure states. The main goal is to
study equivalence classes under general local transformations of
the kind $|\psi'\rangle=A_1\otimes \cdots\otimes A_n|\psi\rangle$
with $\{A_i\}$ arbitrary matrices. These kind of transformations
are called SLOCC transformations \cite{DVC00} (from stochastic
local operations assisted by classical communication), and are
also called local filtering operations. It will turn out very
useful to restrict ourselves to SLOCC transformations where all
$\{A_i\}$ are full rank (remark that entanglement is lost whenever
a $A_i$ is not full rank). For convenience, we will consider all
$\{A_i\}$ to belong to $SL(n,\mathcal{C})$, the group of square
complex matrices having determinant equal to 1, and consider
unnormalized states.

Let us formulate the following central theorem:

\begin{theorem}\label{thnormalform}
Consider an $N_1\times N_2\times \cdots N_p$ pure multipartite
state (or tensor). Then this state (tensor) can constructively be
transformed into a normal form by determinant 1 SLOCC operations.
The local density operators of the normal form are all
proportional to the identity, and the normal form is unique up to
local unitary transformations. Moreover, the state connected to
the original one by determinant 1 SLOCC operations with the
minimal possible norm (i.e. trace of the unnormalized density
operator) is in normal form.
\end{theorem}

{\em Proof:} We will give a constructive proof of this theorem
that can directly be translated into matlab code. The idea is that
the local determinant 1 operators $A_i$ bringing $\psi$ into its
normal form can be iteratively determined by a procedure where at
each step the trace of $|\psi\rangle\langle\psi|=\rho$ is
minimized by  a local filtering operation of one party. Consider
therefore the partial trace $\rho_1={\rm Tr}_{2\cdots p}(\rho)$.
If $\rho_1$ is full rank, there exists an operator $X$ with
determinant 1 such that $\rho_1'=X\rho_1X^\dagger\sim I_{N_1}$.
Indeed, $X=|\det(\rho_1)|^{1/2N_1}(\sqrt{\rho_1})^{-1}$ does the
job\footnote{Note that the numerical algorithm should not
calculate $X$ from $\rho_1$ but instead from the singular value
decomposition of the $N_1\times (\Pi_{i>1}N_i)$ matrix
$\psi_{i_1,(i_2\cdots i_p)}=U\Sigma V^\dagger$: $X$ can be chosen
as $X=\Sigma^{-1}U^\dagger$, and the numerical accuracy will be
much higher.}, and we have $\rho_1'=\det(\rho_1)^{1/N_1}I_{N_1}$.
We also have the relation:
\begin{equation}{\rm Tr}\left(\rho'\right)=N_1\det(\rho_1)^{1/N_1}\leq{\rm Tr}\left(\rho_1\right),\label{decr}\end{equation}
where $\rho'=(X\otimes I\cdots \otimes
I)|\psi\rangle\langle\psi|(X\otimes I\cdots \otimes I)^\dagger$.
This inequality follows from the fact that the geometric mean is
always smaller than the arithmetic mean, with equality iff
$\rho_1$ is proportional to the identity. Therefore the trace of
$\rho$ decreases after this operation. We can now repeat this
procedure with the other parties, and then repeat everything
iteratively over and over again. After each iteration, the trace
of $\rho$ will decrease unless all partial traces are equal to the
identity. Because the trace of a positive definite operator is
bounded from below, we know that the decrements become arbitrarily
small and following equation (\ref{decr}) this implies that all
partial traces converge to operators arbitrarily close to the
identity.\\
We still have to consider the case where we encounter a $\rho_i$
that is not full rank. Then there exists a series of $X$ whose
norm tends to infinity but has determinant 1 such that
$X\rho_iX^\dagger=0$, leading to a normal form identical to zero,
clearly the positive operator with minimal possible trace. This
ends the proof of the existence of the normal form.

Consider now  the a state that is normal form; then due to the
construction  of the proof, the trace can always be decreased by
determinant 1 SLOCC operations, unless the state is in normal
form.

As pointed out by Briand, Luque and Thibon \cite{BLT}, the normal
form is unique up to local unitaries: the Kempf-Ness criterion
proves the uniqueness in the case of a closed orbit, and there is
always a unique closed orbit in the closure of an arbitrary orbit
\cite{Scha}. This ends the proof.\qed

Let us now return to the general theorem \ref{thnormalform}.  This
theorem is very fundamental in that it states that each pure
multipartite state can be transformed into a unique state  with
the property that all local density operators are proportional to
the identity. States in normal form are clearly expected to be
maximally entangled states. As we will argue later, the normal
form is the state with the maximal amount of entanglement that can
be created locally and probabilistically from the original state.

Let us next prove that the normal form is continuous with respect
to perturbations of the entries of the original density matrix
$\rho$. First of all note that the non-uniqueness due to the local
unitaries can be circumvented by imposing all $A_i$ to be
hermitian. The following lemma shows that the normal form is
robust against perturbations or noise:
\begin{lemma}
If the SLOCC operations bringing the state into the normal form
introduced in theorem \ref{thnormalform} are chosen to be
hermitian, and if they turn out to be finite,  then the normal
form is continuous with respect to the entries of the state.
\end{lemma}
Proof:  Let us consider $\rho=(A_1\otimes\cdots \otimes
A_p)\sigma(A_1\otimes\cdots \otimes A_p)^\dagger$ and a
perturbation $\dot{\rho}$ resulting in $\{\dot{A_i}\}$ and
$\dot{\sigma}$. The following formula is readily verified:
\bea{&&(A_1\otimes\cdots \otimes
A_p)^{-1}\dot{\rho}(A_1\otimes\cdots \otimes
A_p)^{-\dagger}=\nonumber\\&&\hspace{.5cm}\dot{\sigma}+\sum_{i=1}^p
\left((I\otimes \cdots A_i^{-1}\dot{A_i}\cdots\otimes
I)\sigma+{\rm h.c.}\right).\nonumber} As all $\{A_i\}$ are
hermitian and have determinant 1, all $A_i^{-1}\dot{A_i}$ are
skew-hermitian and the second term lives in another subspace $S_2$
than the first term $\dot{\sigma}$ who lives in subspace $S_1$.
$\dot{\sigma}$ can therefore be obtained by projecting the left
hand side parallel to $S_2$ onto $S_1$. As $\dot{\rho}$ is finite
and all $\{A_i\}$ have determinant one and are finite, this
projection is of course also finite. This proofs that
$\dot{\sigma}$ is of the same order of magnitude as $\dot{\rho}$,
which ends the proof.\qed

Note that we have also proven continuity with respect to mixing.

Let us now discuss some peculiarities. The fact that the algorithm
can converge to zero despite the fact that all $A_i$ have
determinant equal to $1$ is a consequence of the fact that
$SL(n,\mathcal{C})$ is not compact: there exist states that can
only be brought into their respective normal form by infinite
transformations, although the class of states with this property
is clearly of measure zero. As an example consider the $W$-state
\cite{DVC00} $|\psi\rangle=|001\rangle+|010\rangle+|100\rangle$.
The following identity is easily checked:
\[\lim_{t\rightarrow\infty} \left(\begin{array}{cc} 1/t&0\\0&t
\end{array}\right)^{\otimes 3}|W\rangle=0.\] The normal form
corresponding to the $W$-state is therefore equal to zero, clearly
the state with the minimal possible trace. This is interesting, as
it will be shown that a normal form is zero iff a whole class of
entanglement monotones is equal to zero. Therefore the states with
normal form equal to zero are fundamentally different from those
with finite normal form, and this leads to the generalization of
the $W$-class to arbitrary dimensions.

It thus happens that some states have normal form equal to 0. This
also happens if the state does not have full support on the
Hilbert space in that one partial trace $\rho_i$ is rank
deficient. Note that states which do not have full support on the
Hilbert space, such as pure states from which one party is fully
separable, all have normal form equal to zero. It will indeed turn
out that the amount of multipartite entanglement present in a
state can be quantified by the trace of the obtained normal form,
which is clearly zero in the case of separable states. On the
other hand, the only normalized states that are already in normal
form are precisely the maximally entangled states:  in the case of
three qubits for example, the only state with the property that
all its local density operators are proportional to the identity
is the GHZ-state.

As a last remark,  we give an example of a state that is brought
into a non-zero normal form by SLOCC operators that are unbounded:
\be{|\psi\rangle\simeq
a(|0000\rangle+|1111\rangle)+|01\rangle(|10\rangle+|01\rangle)}
The normal form is just given by the GHZ-state
$(|0000\rangle+|1111\rangle)$, but as can be derived from the
results presented  in \cite{VDD02b}, infinite SLOCC
transformations are needed to reach this.

\subsection{Entanglement monotones}\label{EMgen}
Until now we contented ourselves to characterize the orbits
generated by local unitary or SLOCC operations, but we have not
tried to quantify the entanglement present in a state. The SLOCC
normal form introduced in the previous section however gives us a
strong hint of how to do this. Note that all separable states have
a  normal form equal to zero, and that the known maximally
entangled states such as Bell-states and GHZ-states are the only
ones of their class that are in normal form.

This suggests a very general way of constructing entanglement
monotones:
\begin{theorem}
Consider a linearly homogeneous positive function of a pure
(unnormalized) state $M(\rho=|\psi\rangle\langle\psi|)$ that
remains invariant under determinant 1 SLOCC operations. Then
$M(|\psi\rangle\langle\psi|)$ is an entanglement monotone.
\label{thmon}
\end{theorem}
{\em Proof:} A quantity $M(\rho)$ is an entanglement monotone iff
its expected value does not increase  under the action of every
local operation. It is therefore sufficient to show that for every
local $A_1\leq I_{N_1}$, $\bar{A}_1=\sqrt{I_{N_1}-A_1^\dagger
A_1}$, it holds that
\begin{eqnarray}M(\rho)\geq &&{\rm
Tr}\left((A_1\otimes I)\rho(A_1\otimes
I)^\dagger\right).\nonumber\\&&\hspace{1cm}M\left(\frac{(A_1\otimes
I)\rho(A_1\otimes I)^\dagger}{{\rm Tr}\left((A_1\otimes
I)\rho(A_1\otimes
I)^\dagger\right)}\right)\nonumber\\
&&+{\rm Tr}\left((\bar{A_1}\otimes I)\rho(\bar{A_1}\otimes
I)^\dagger\right).\nonumber\\&&\hspace{1cm}M\left(\frac{(\bar{A_1}\otimes
I)\rho(\bar{A_1}\otimes I)^\dagger}{{\rm
Tr}\left((\bar{A_1}\otimes I)\rho(\bar{A_1}\otimes
I)^\dagger\right)}\right)\nonumber\end{eqnarray}

If $A_1$ is full rank, it can be transformed to a determinant 1
matrix by dividing it by $\det(A_1)^{1/N_1}$. Due to the
homogeneity of $M(\alpha\rho)=\alpha M(\rho)$  the previous
inequality is equivalent to
\[M(\rho)\geq
(|\det(A_1)|^{2/N_1}+|\det(\bar{A_1})|^{2/N_1})M(\rho).\] As the
arithmetic mean always exceeds the geometric mean,  this
inequality is always satisfied. This argument can be easily
completed to the cases where $A_i$ is not full rank due to
continuity. The same argument can then be repeated for the other
$A_i$, which ends the proof.\qed

Entanglement monotones of the above class can readily be
constructed using the completely antisymmetric tensor
$\epsilon_{i_1\cdots i_N}$.\\
Indeed, it holds that $\sum A_{i_1j_1}A_{i_2j_2}\cdots
A_{i_Nj_N}\epsilon_{j_1\cdots j_N}=\det(A)\epsilon_{i_1\cdots
i_N}$, and as $\det(A)=1$ this leads to invariant quantities under
determinant 1 SLOCC operations. These quantities seem to be
related to hyperdeterminants \cite{GKZ94,Miy02}, and those latter
seem to be a subclass of the quantities considered here.

Consider for example the case of two qubits. The quantity
\[|\sum_{i_1j_1i_2j_2}
\psi_{i_1j_1}\psi_{i_2j_2}\epsilon_{i_1i_2}\epsilon_{j_1j_2}|\] is
clearly of the considered class, and it happens to be the
celebrated concurrence  entanglement measure \cite{Woo98}. In the
case of three qubits, the simplest non-trivial homogeneous
quantity invariant under determinant 1 SLOCC operations is given
by
\[|\psi_{i_1j_1k_1}\psi_{i_2j_2k_2}\psi_{i_3j_3k_3}\psi_{i_4j_4k_4}\epsilon_{i_1i_2}\epsilon_{i_3i_4}\epsilon_{j_1j_2}\epsilon_{j_3j_4}\epsilon_{k_1k_3}\epsilon_{k_2k_4}|^{1/2}\]
(Note that we use the Einstein summation convention.) This happens
to the square root of the 3-tangle introduced by Wootters et
al.\cite{CKW00}, which quantifies the true tripartite
entanglement.

More generally, as the considered entanglement monotones are
invariant under the determinant 1 SLOCC operations, the number of
independent entanglement monotones is equal to the degrees of
freedom of the normal form obtained in the case of a pure state
minus the degrees of freedom induced by the local unitary
operations. Indeed, this is the amount of invariants of the whole
class of states connected by SLOCC operations. It is then easily
proven that a normal form is equal to zero if and only if all the
considered entanglement monotones are equal to zero: the
entanglement monotones are homogeneous functions of the normal
form, and if the normal form is not equal to zero there always
exists an SLOCC invariant quantity that is different from zero.

In the case of 4 qubits for example, parameter counting leads to
$(2\cdot 2^4-2)-4\cdot 6=6$ (a state has $32$ degrees of freedom
$-2$ to an irrelevant phase and the $4$ $SL(2,\mathcal{C})$
matrices have each $6$ degrees of freedom) independent
entanglement monotones. The simplest monotone is given by
\be{|\psi_{i_1j_1k_1l_1}\psi_{i_2j_2k_2l_2}\epsilon_{i_1i_2}\epsilon_{j_1j_2}\epsilon_{k_1k_2}\epsilon_{l_1l_2}|,\label{tangle2222a}}
and the other 5 entanglement monotones can be obtained by
including more factors; an example is
\begin{eqnarray}
&&\sqrt{2}|\psi_{i_1j_1k_1l_1}\psi_{i_2j_2k_2l_2}\psi_{i_3j_3k_3l_3}\psi_{i_4j_4k_4l_4}.\nonumber\\
&&\hspace{1cm}\epsilon_{i_1i_2}\epsilon_{i_3i_4}\epsilon_{l_1l_2}\epsilon_{l_3l_4}\epsilon_{j_1j_3}\epsilon_{j_2j_4}\epsilon_{k_1k_3}\epsilon_{k_2k_4}|^{1/2}\label{tangle2222b}\end{eqnarray}
These are clearly generalizations of the concurrence and the
3-tangle to four parties. Note however that the situation here is
more complicated due to the existence of multiple independent
entanglement monotones. Note also that there exist biseparable
states that can be brought into a non-zero normal form by
determinant 1 SLOCC operations. Consider for example the tensor
product of two Bell states; all local density operators are
proportional to the identity, the value of the entanglement
monotones (\ref{tangle2222a}) and (\ref{tangle2222b}) is
respectively given by $1$ and $1/\sqrt{2}$ (as opposed to $1$ and
$1$ for the GHZ-state $(|0000\rangle+|1111\rangle)/\sqrt{2}$), and
nevertheless no true 4-partite entanglement is present.

If the subsystems happen to be of unequal dimension, then the
respective subdimensions should be chosen not larger than the
maximal allowed dimension such that all local density operators
remain full rank. In a $2\times 2\times N$ system for example, a
pure state can only have full support on the $2\times 2\times 4$
subspace, and therefore it makes no sense to calculate the normal
form with $N>4$: one can always first rotate the $N$-dimensional
system into a $4$-dimensional one by local unitary operations, and
proceed by calculating the normal form for the $2\times 2\times 4$
system. More generally, if the dimension of the largest subsystem
does not exceed the product of all the other ones, then
generically the normal form will not be equal to zero, leading to
non-trivial entanglement monotones. As an example, consider a
$2\times 2\times 4$ system; there are more local SLOCC parameters
than the number of degrees of freedom, so there will be only one
entanglement monotone (as is the case in the $2\times 2$ and
$2\times 2\times 2$ case). The $2\times 2\times 4$ tangle is given
by: \begin{eqnarray*}&&\sqrt{4/3}|\sum
\psi_{i_1j_1k_1}\psi_{i_2j_2k_2}\psi_{i_3j_3k_3}\psi_{i_4j_4k_4}.\\&&\hspace{3cm}\epsilon_{i_1i_2}\epsilon_{i_3i_4}\epsilon_{j_1j_3}\epsilon_{j_2j_4}\epsilon_{k_1k_2k_3k_4}|^{1/2}\end{eqnarray*}
The factor $\sqrt{4/3}$ is included to ensure that the state in
normal form
\be{(|000\rangle+|011\rangle+|102\rangle+|113\rangle)/2} has
tangle given by $1$. Indeed, as will be shown in the following
section, the maximal value of the tangle is always obtained for
states in normal form, and this is the unique state  (up to LU)
having all its local density operators proportional to the
identity. Note that this state is therefore the generalization of
the $GHZ$ state to $2\times 2\times 4$ systems.

For completeness, let us also give a formula for the $2\times
2\times 3$ tangle:
\begin{eqnarray*}&&\hspace{-.5cm}\sqrt[3]{\frac{27}{4}}|\sum
\psi_{i_1,j_1,k_1}\psi_{i_2,j_2,k_2}\psi_{i_3,j_3,k_3}\psi_{i_4,j_4,k_4}\psi_{i_5,j_5,k_5}\psi_{i_6,j_6,k_6}.\\
&&\hspace{1.7cm}
\epsilon_{i_1i_4}\epsilon_{i_2i_5}\epsilon_{i_3i_6}\epsilon_{j_1j_4}\epsilon_{j_2j_5}\epsilon_{j_3j_6}
\epsilon_{k_1k_2k_3}\epsilon_{k_4k_5k_6}|^{1/3}\end{eqnarray*} The
state maximizing this entanglement monotone (the number is bounded
by $1$) is the generalization of the $GHZ$ to the $2\times 2\times
3$ case:
\be{\frac{1}{\sqrt{3}}|000\rangle+\frac{1}{\sqrt{6}}|011\rangle+\frac{1}{\sqrt{6}}|101\rangle+\frac{1}{\sqrt{3}}|112\rangle.}

Let us finally give a non-trivial example of an entanglement
monotone of the considered class in the case of three qutrits:
\begin{eqnarray*}&&\sqrt{2}|\sum\psi_{i_1 j_1 k_1}\psi_{i_2 j_2 k_2}\psi_{i_3
j_3 k_3}\psi_{i_4 j_4 k_4}\psi_{i_5 j_5 k_5}\psi_{i_6 j_6
k_6}\nonumber\\
&&\hspace{1.5cm}\epsilon_{i_1 i_2 i_3}\epsilon_{i_4 i_5
i_6}\epsilon_{j_1 j_2 j_4}\epsilon_{j_3 j_5 j_6} \epsilon_{k_1 k_5
k_6}\epsilon_{k_2 k_3 k_4}|^{1/3}.\end{eqnarray*} The other
$(2\cdot 3^3-1)-(3\cdot 16)-1=4$ independent entanglement
monotones can again be constructed by including more factors.

\subsection{Optimal Filtering}

A natural question now arises:  characterize the optimal SLOCC
operations to be performed on one copy of a multipartite system
such that, with a non zero chance, a state with maximal possible
multipartite entanglement is obtained. This question is of
importance for experimentalists as in general they are not able to
perform joint operations on multiple copies of the system.
Therefore the procedure outlined here often represents the best
entanglement distillation procedure that is practically
achievable.

In the previous section a whole class of entanglement monotones
that measures the amount of multipartite entanglement were
introduced. The following theorem can easily be proved using the
techniques of theorem \ref{thnormalform}
\begin{theorem}\label{thoptfilt}
Consider a pure multipartite state, then the local filtering
operations that maximize all entanglement monotones introduced in
theorem \ref{thmon} are represented by operators proportional to
the determinant 1 SLOCC operations that transform the state into
its normal form.
\end{theorem}
{\em Proof:} The proof of this theorem is surprisingly simple.
Indeed, all the quantities introduced in theorem \ref{thmon} are
invariant under determinant 1 SLOCC operations if the states do
not get normalized. The value of an entanglement monotone however
only makes sense if defined on normalized states, and due to the
linear homogeneity of the entanglement monotones, the following
identity holds:
\[M\left(\frac{(\otimes_i A_i)\rho(\otimes_i
A_i)^\dagger}{{\rm Tr}\left((\otimes_i A_i)\rho(\otimes_i
A_i)^\dagger\right)}\right)=\frac{M(\rho)}{{\rm
Tr}\left((\otimes_i A_i)\rho(\otimes_i A_i)^\dagger\right)}\] The
optimal filtering operators are then obtained by the $\{A_i\}$
minimizing \be{{\rm Tr}\left((\otimes_i A_i)\rho(\otimes_i
A_i)^\dagger\right).} But this problem was solved in theorem
\ref{thnormalform}, where it was proved that the $\{A_i\}$
bringing the state into its normal form minimize this trace.\qed

It is therefore proved that the (reversible) procedure of washing
out the local correlations maximizes the multipartite entanglement
as measured by the generalization of the tangle. This is in
complete accordance with the results of majorization \cite{Nie99},
where it is shown that the notion of local disorder is intimately
connected to the amount of entanglement present. Therefore we have
supporting evidence to call pure states in normal form maximally
entangled with relation to their SLOCC orbit.

\subsection{The mixed state case.}
The normal form derived in theorem \ref{thnormalform} can readily
be generalized to the case where the state is mixed, i.e. the case
where the density operator is a convex sum of pure states. Indeed,
nowhere in the proof of the theorem it was used that the state
$\rho$ was pure; the same holds for the continuity for the normal
form. We have therefore proven:

\begin{theorem}
Consider an $N_1\times N_2\times \cdots N_m$ mixed multipartite
state. Then this state can be brought into a normal form by
determinant 1 SLOCC operations, where the normal form has all
local density operators proportional to the identity, and the
normal form is unique up to local unitary operations. Moreover the
trace of the normal form is the minimal one that can be obtained
by determinant 1 SLOCC operations. If the SLOCC operations are
chosen to be hermitian, then the normal form is continuous with
respect to perturbations of the original state.
\end{theorem}

Note that if $\rho$ is full rank, its normal form will never
converge to zero: the determinant of the density operator is
constant under SLOCC operations.

It is also possible to adopt the results about entanglement
monotones. First of all we extend the definition of an
entanglement monotone $\mu_p$ that is defined on pure states and
that is linearly homogeneous in $\rho$ by the convex roof
formalism:
\be{\mu_m(\rho)=\min_{\sum_ip_i|\psi_i\rangle\langle\psi_i|=\rho}\sum_ip_i\mu_p(|\psi_i\rangle).}
Here the optimization has to be done over all pure state
decompositions of the state. The fact that the pure state
entanglement monotone is linearly homogeneous in $\rho$ ensures
that $\mu_m$ is, on average, not increasing under local
operations, and therefore assures that $\mu_m$ is an entanglement
monotone. Moreover, it is obvious that these entanglement
monotones are again invariant under determinant $1$ SLOCC
operations. The results on optimal filtering for mixed states also
readily apply, and  therefore we arrive at the following  very
powerful result:
\begin{theorem}\label{optfilmixed}
The local filtering operations  bringing a mixed state into its
normal form are exactly the ones that maximize the entanglement
monotones that remain invariant under determinant 1 SLOCC
operations.
\end{theorem}

This result is remarkable, because there does typically not exist
a way of actually calculating the value of an entanglement
monotone defined on a mixed state: finding the optimal pure state
decomposition of a state with relation to the convex roof
formalism for a given  EM is excessively difficult and has until
now only been proven possible for the concurrence (i.e. the case
of two qubits). So although we cannot calculate the entanglement
monotone, we know how to maximize it! This particularly applies to
mixed states of three qubits: we have proven how to maximize the
3-tangle, although we don't know how to calculate it.

Note that this optimal filtering procedure produces non-trivial
results even in the case of two qubits: it proves that the
concurrence and therefore the entanglement of formation of a mixed
state of two qubits is  maximized by the SLOCC operations bringing
the state into its unique (Bell-diagonal) normal form.

\subsection{Conclusion}
In conclusion, we presented a constructive way of bringing a
single copy of a quantum state into normal form under local
filtering operations. This normal form is such that all {\em
local} information is washed out (i.e. the local density operators
are maximally mixed). We presented qualitative and quantitative
arguments why the amount of entanglement of states in normal form
cannot be enlarged by local operations, and introduced a whole
class of entanglement measures that are a direct generalization of
concurrence and 3-tangle to systems of arbitrary dimension. This
sheds some new light on the difficulty encountered in classifying,
understanding and unravelling the mysteries of multipartite
quantum entanglement.

\acknowledgements We are very grateful to E. Briand, J-G. Luque
and J-Y. Thibon for pointing out the theorems in algebraic
geometry that prove the uniqueness of the normal form.

\appendix
\section{Appendix: Normal forms under local unitary operations}

Consider a general multipartite state with $m$ parties defined on
a $n_1\otimes n_2\cdots n_M$ dimensional Hilbert space:
\be{|\psi\rangle=\sum_{i_1\cdots i_m}\psi_{i_1\cdots
i_m}|i_1\rangle|i_2\rangle\cdots|i_m\rangle.}

In this appendix, we try to solve the following  natural question:
is there a method to verify if two states $|\psi_1\rangle$ and
$|\psi_2\rangle$ are equivalent up to local unitary
transformations? In the bipartite case, this problem can readily
be solved using the singular value decomposition, and we therefore
ask for some kind of generalization of this diagonal normal form.
Let us state the following theorem (see also Carteret et al.
\cite{CHS00}), which is a weak generalization of the SVD:

\begin{theorem}\label{thnormalUgen}
Given a general complex tensor $\psi_{i_1\cdots i_m}$ with
dimensions $n_1= n_2=\cdots= n_m=n$, then there exist local
unitaries $U_i$ such that all the following entries in the tensor
$\psi'=U_1\otimes \cdots\otimes U_m\psi_{i_1\cdots i_m}$ are set
equal to zero:
\begin{eqnarray*}\forall 1\leq j\leq n, \forall
k>j:&& \psi'_{j,j,\cdots,j,j,k}=0\\
&&\psi'_{j,j,\cdots j,k,j}=0\\
&&\vdots\\
&&\psi'_{j,k,j,\cdots ,j,j}=0\\
&&\psi'_{k,j,\cdots ,j,j}=0.\end{eqnarray*} Moreover all entries
$\psi'_{n,n,\cdots,n,i,n,\cdots n}, i\leq n$ can be made real and
positive. If the number of parties exceeds 2, then the normal form
is typically not unique up to permutations, but there exist a
discrete number of different normal forms with the aforementioned
property.  The number of zeros however can generically not  be
increased by further local unitary operations.
\end{theorem}
Proof: unlike the proof in \cite{CHS00}, this proof is
constructive and can readily be translated into  matlab code to
calculate the normal form numerically. Consider first all entries
with at least $m-1$ times $1$ in its indices, and define the
vectors $x^1_i=\psi_{i,1,1,\cdots,1}$,
$x^2_i=\psi_{1,i,1,\cdots,1}$, $\dots$ ,
$x^m_i=\psi_{1,1,\cdots,1,i}$. Define now a recursive algorithm
that goes as follows: rotate $x^1$ to $\|x^1\|[1,0,\cdots 0]$ by a
unitary transformation, apply the same transformtion on the full
tensor, and define $x^2=\psi_{1,i,1,\cdots,1}$ with $\psi$ the
transformed tensor. Now do the same thing with $x^2$, \dots $x^m$
and then again with $x^1$, until the algorithm converges. This
algorithm will certainly converge because at each step the
$(1,1,\cdots 1)$ entry of $\psi$ becomes larger and larger, unless
all entries $(1,1,\cdots, 1,i,1,\cdots 1)$ are equal to zero;
moreover its value is bounded above because the unitary group is
compact. Next exactly the same algorithm can be applied to the
subtensor of $\psi$ defined as the one with all entries larger or
equal to 2 (it is easy to check that the zeros obtained in the
first step will remain zero by this kind of action). Next we can
again do the same thing of another (smaller) subtensor, proving
that indeed all zeros quoted in the theorem can
be made.\\
It is straightforward to prove that the entries
$\psi'_{n,n,\cdots,n,i,n,\cdots n}, i\leq n$ can all be made real
and positive by further diagonal unitary transformations.\\
Let us finally prove that no more zeros can be made by whatever
unitaries (in the generic case). This follows from the fact that a
unitary $n\times n$ matrix has $n^2$ continuous real degrees of
freedom, but that only $n^2-n$ of them can be used to produce
zeros as the other $n$ degrees of freedom can be imbedded in a
diagonal unitary with just phases. Counting of the number of zeros
produced indeed leads to
\be{\sum_{j=1}^m\sum_{k=1}^{m-1}\max(n-k,0)= m\frac{n(n-1)}{2}}
which indeed corresponds to the $m(n^2-n)$
degrees of freedom as the zeros are "complex".\\
The non-uniqueness of the normal form obtained is surprising but
can readily be verified by implementing the algorithm on a generic
tensor; typically the algorithm converges to one out of a finite
number of possible different normal forms.\qed

As a first example, consider a system of three qubits. Unfolding
the $2\times 2\times 2$ tensor in two $2\times 2$ matrices, the
following entries can always be made equal to zero:
\be{\ba{cc}{\ba{cc}{x&0\\0&x}\ba{cc}{0&x\\x&x}}} Here $x$ is used
to denote a non-zero entry. In this case, it is easy to see that 4
of the remaining 5 entries can be made real by multiplying with
appropriate diagonal local unitaries. This is equivalent to the
normal form obtained by Acin et al.\cite{AAC00}.

A more sophisticated example is the $3\times 3\times 3$ case,
whose normal form looks like
\be{\ba{ccc}{\ba{ccc}{x&0&0\\0&x&x\\0&x&x}\ba{ccc}{0&x&x\\x&x&0\\x&0&x}\ba{ccc}{0&x&x\\x&0&x\\x&x&x}}\label{333}}
It is also straightforward to generalize the previous theorem (and
constructive proof) to systems with different subdimensions (see
Carteret et al.\cite{CHS00} for an existence proof); the algorithm
of the previous proof can readily be extended to this case. Let us
for example consider the normal form of the $N\times 2\times 2$
case:
\be{\ba{cc}{\ba{cc}{x&0\\0&x\\0&x\\0&x\\0&0\\\vdots&\vdots\\0&0}\ba{cc}{0&x\\x&x\\x&0\\0&0\\0&0\\\vdots&\vdots\\0&0}}}
This case is of particular interest as it is describes a state of
two qubits entangled with the rest of the world.

\bibliographystyle{unsrt}

\begin{thebibliography}{10}

\bibitem{BBP96a}
C.H. Bennett, H.J. Bernstein, S.~Popescu, and B.~Schumacher.
\newblock Concentrating partial entanglement by local operations.
\newblock {\em Phys. Rev. A}, 53:2046--2052, 1996.

\bibitem{Gi96}
N.~Gisin.
\newblock Hidden quantum nonlocality revealed by local filters.
\newblock {\em Phys. Lett. A}, 210:151, 1996.

\bibitem{HHH97}
M.~Horodecki, P.~Horodecki, and R.~Horodecki.
\newblock Inseparable two spin 1/2 density matrices can be distilled to a
  singlet form.
\newblock {\em Phys. Rev. Lett.}, 78:574--577, 1997.

\bibitem{LMP98}
N.~Linden, S.~Massar, and S.~Popescu.
\newblock Purifying noisy entanglement requires collective measurements.
\newblock {\em Phys. Rev. Lett.}, 81:3279, 1998.

\bibitem{Ken98}
A.~Kent.
\newblock Entangled mixed states and local purification.
\newblock {\em Phys. Rev. Lett.}, 81:2839, 1998.

\bibitem{KLM99}
A.~Kent, N.~Linden, and S.~Massar.
\newblock Optimal entanglement enhancement for mixed states.
\newblock {\em Phys. Rev. Lett.}, 83:2656, 1999.

\bibitem{VDD01b}
F.~Verstraete, J.~Dehaene, and B.~De Moor.
\newblock Local filtering operations on two qubits.
\newblock {\em Phys. Rev. A}, 64:010101(R), 2001.

\bibitem{DVC00}
W.~D{\" u}r, G.~Vidal, and J.I. Cirac.
\newblock Three qubits can be entangled in two inequivalent ways.
\newblock {\em Phys. Rev. A}, 62:062314, 2000.

\bibitem{Vid00}
G.~Vidal.
\newblock Entanglement monotones.
\newblock {\em Jour. of Modern Optics}, 47(2/3):355--376, 2000.

\bibitem{CKW00}
V.~Coffman, J.~Kundu, and W.K. Wootters.
\newblock Distributed entanglement.
\newblock {\em Phys. Rev. A}, 61:052306, 2000.

\bibitem{Woo98}
W.K. Wootters.
\newblock Entanglement of formation of an arbitrary state of two qubits.
\newblock {\em Phys. Rev. Lett.}, 80:2245, 1998.

\bibitem{Nie99}
M.A. Nielsen.
\newblock Conditions for a class of entanglement transformations.
\newblock {\em Phys. Rev. Lett.}, 83:436--439, 1999.


\bibitem{BLT} E. Briand, J-G. Luque and J-Y. Thibon, e-print quant-ph/0304026.

\bibitem{Scha} I.R. Shafarevich.
\newblock {\em Algebraic geometry IV}
\newblock  Vol. 55 of Encycopedia of Mathematical Sciences;
\newblock Springer-Verlag, Berlin, 1994.


\bibitem{VDD02b}
F.~Verstraete, J.~Dehaene, B.~De Moor, and H.~Verschelde.
\newblock Four qubits can be entangled in nine different ways.
\newblock {\em Phys. Rev. A}, 65:052112, 2002.

\bibitem{Miy02}
A.~Miyake.
\newblock Topological classification of multipartite entangled states by the
  hyperdeterminant.
\newblock 2002.
\newblock quant-ph/0206111.

\bibitem{CHS00}
H.A. Carteret, A.~Higuchi, and A.~Sudbery.
\newblock Multipartite generalization of the schmidt decomposition.
\newblock {\em J. Math. Phys.}, 41:7932, 2000.

\bibitem{AAC00}
A.~Ac\'{\i}n, A.~Andrianov, L.~Costa, E.~Jan\'{e}, J.~I. Latorre,
and
  R.~Tarrach.
\newblock Schmidt decomposition and classification of three-quantum-bit states.
\newblock {\em Phys. Rev. Lett.}, 85:1560, 2000.

\end{thebibliography}

\end{document}